\documentclass[twocolumn,showpacs,superscriptaddress,prl]{revtex4}
\pdfoutput=1
\usepackage{amsmath, amsthm, amssymb}
\usepackage{graphicx}
\usepackage{dcolumn}
\usepackage{bm}
\usepackage{color} %for change tracking

\newcommand{\ket}[1]{| #1 \rangle}
\newcommand{\bra}[1]{\langle #1 |}

\newcommand{\ie}{{\it{i.e.~}}}
\newcommand{\etal}{{\it{et al.}}}

\begin{document}

\title{Open-system dynamics of graph-state entanglement}
\author{Daniel Cavalcanti}
\affiliation{ICFO-Institut de Ciencies Fotoniques, Mediterranean
Technology Park, 08860 Castelldefels (Barcelona), Spain}
\author{Rafael Chaves}
\affiliation{Instituto de F\'\i sica, Universidade Federal do Rio de Janeiro. Caixa Postal
68528, 21941-972 Rio de Janeiro, RJ, Brasil}
\affiliation{Physikalisches Institut der Albert-Ludwigs-Universitat,
Hermann-Herder-Str. 3, D-79104 Freiburg, Germany}
\author{Leandro Aolita}
\affiliation{ICFO-Institut de Ciencies Fotoniques, Mediterranean
Technology Park, 08860 Castelldefels (Barcelona), Spain}
\author{Luiz Davidovich}
\affiliation{Instituto de F\'\i sica, Universidade Federal do Rio de Janeiro. Caixa Postal
68528, 21941-972 Rio de Janeiro, RJ, Brasil}
\author{Antonio Ac\'in}
\email{antonio.acin@icfo.es}
\affiliation{ICFO-Institut de Ciencies Fotoniques, Mediterranean
Technology Park, 08860 Castelldefels (Barcelona), Spain}
\affiliation{ICREA-Instituci\'o Catalana de Recerca i Estudis
Avan\c cats, Lluis Companys 23, 08010 Barcelona, Spain}

\begin{abstract}
We consider graph states of arbitrary number of particles
undergoing generic decoherence. We present methods to obtain lower
and upper bounds for the system's entanglement in terms of that of
considerably smaller subsystems. For an important class of noisy
channels, namely the Pauli maps, these bounds coincide
and thus provide the exact analytical expression for the
entanglement evolution. All the results apply also to (mixed) graph-diagonal states, and hold true for
any convex entanglement monotone. Since any state can be
locally depolarized to some graph-diagonal state, our method provides a lower bound
for the entanglement decay of any arbitrary state.
Finally, this formalism also allows for the direct identification of the robustness under size scaling of graph states in the presence of decoherence, merely by inspection of their connectivities.
\end{abstract}


\maketitle

\emph{Introduction.--} Graph states~\cite{graph_review} constitute an
important class of entangled states with broad-reaching
applications in quantum information, including measurement-based
quantum computation~\cite{Brie_review,RausBrie},  quantum error
correction~\cite{SchWer}, and secure quantum communication~\cite{DurCasBrie-ChenLo}. Moreover, instances of this family, such as the Greenberger-Horne-Zeilinger
states, play a crucial role in fundamental tests of quantum non-locality~\cite{GHZ}.  Consequently, a
great effort has been made both to theoretically understand their
properties~\cite{graph_review,HeinEisBrie} and to create
and coherently manipulate them experimentally~\cite{ClusterExp}

Needless to say, it is crucial to understand the dynamics of their
entanglement  in realistic scenarios, where the system unavoidably
decoheres due to experimental errors or to the interaction with
its environment. Previous studies on the robustness of graph-state
entanglement in the presence of decoherence observed a disentanglement time (or lower bounds thereof) insensitive to the system
size \cite{Simon&Kempe, HeinDurBrie}. However,  the disentanglement time on its own is not in general able to provide
any faithful assessment about the entanglement's  robustness, since
it can grow with the number $N$ of particles and yet the
entanglement can get closer to zero the faster, the larger $N$
%robustness since
%it can grow with the number $N$ of particles and yet the amount of
%entanglement go to zero with $N$
\cite{us}. The full dynamical evolution of entanglement must then
be studied to draw conclusions on its fragility. 
Taking the latter
into account, the entanglement of the linear-cluster
states, an example of graph states, was shown to be robust with
the size of the system against the particular case of collective
dephasing decoherence \cite{Guhne}.

The present work provides a general framework for the
study of the entanglement evolution of graph states under decoherence.
Our techniques apply to (i) any graph, and 
graph-diagonal, states; (ii) arbitrary kinds of noise,
individual or collective; and (iii) any convex (bi-
or multi-partite) entanglement quantifier that does not increase
under local operations and classical communication (LOCC). In the developed formalism we consider local
measurement protocols to efficiently obtain lower and upper bounds
for the entanglement of the whole system contained in any given
partition in terms of that of a considerably smaller subsystem consisting only of those qubits lying on the boundary of the partition. No optimization on the full system's parameter space is required
throughout. For an important class of noisy
channels -- namely arbitrary Pauli maps, to be defined below --  the lower and upper bounds coincide, providing thus the exact entanglement
evolution. With the same methods we also establish a
second family of lower bounds that, despite less tight, depend
only on the connectivity of the graph and not on its size. This
allows us to assess the robustness based on the full dynamics of
the entanglement and not just its  disentanglement time.  Our approach can also
be used to establish lower bounds to the entanglement behavior of
any initial quantum state.

%%%%%%DEFYNING GRAPH STATES:

\emph{Graph states.--} Consider a mathematical graph
$G_{(\mathcal{V},\mathcal{E})}\equiv\{\mathcal{V},\mathcal{E}\}$,
composed of a set $\mathcal{V}$, of $N$ vertices $i \in
\mathcal{V}$, and a set $\mathcal{E}$, of  edges  $\{i,j\}\in
\mathcal{E}$ connecting each vertex $i$ to some other $j$. The
associated  physical state is operationally defined as follows: to
each vertex $i$ associate a qubit, initialize all $N$ qubits in
the product state $\ket{{g_{(\mathcal{V})}}_0}\equiv\bigotimes_{i
\in\mathcal{V}} \ket{+_i}$, being
$\ket{+_i}=(\ket{0_i}+\ket{1_i})/\sqrt{2}$,  and to all pairs
$\{i,j\}$ of qubits  joined by an edge apply a
maximally-entangling control-$Z$ ($CZ$) gate,
$CZ_{ij}=\ket{0_{i}0_{j}}\bra{0_{i}0_{j}}+\ket{0_{i}1_{j}}\bra{0_{i}1_{j}}+\ket{1_{i}0_{j}}\bra{1_{i}0_{j}}-\ket{1_{i}1_{j}}\bra{1_{i}1_{j}}$.
The resulting $N$-qubit graph state is
\begin{equation}
\ket{{G_{(\mathcal{V},\mathcal{E})}}_0}=\bigotimes_{\{i,j\} \in \mathcal{E}} CZ_{ij}\ket{{g_{(\mathcal{V})}}_0}.
\end{equation}

\begin{figure}%[t]
\begin{center}
\includegraphics[width=0.7\linewidth]{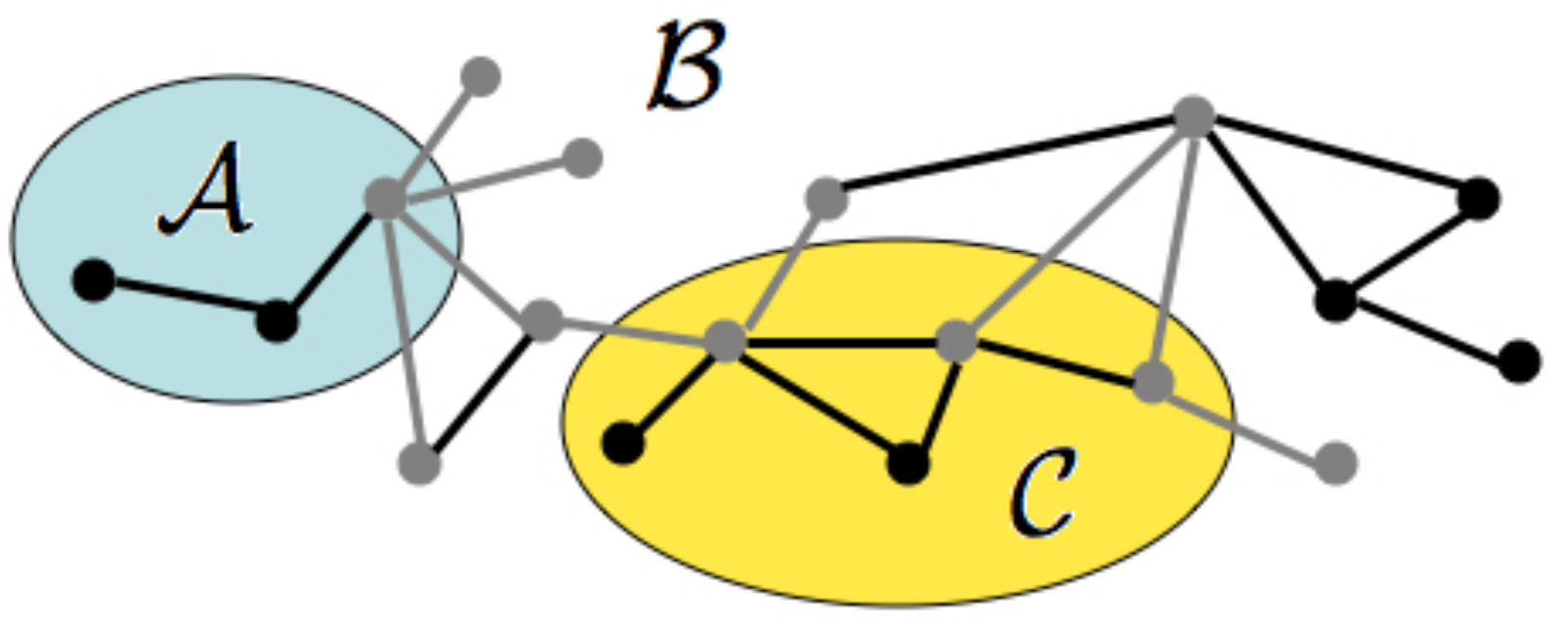}
\caption{\label{Graph} (Color online) Example of a mathematical
graph associated to a physical graph state. We have displayed a
possible partition of this graph, splitting the system in three
parts $\mathcal{A}$, $\mathcal{B}$, and $\mathcal{C}$. The
vertices and edges in grey corresponds to the \emph{boundary
qubits} and the  \emph{boundary-crossing edges} respectively.}
\end{center}
\end{figure}

An example of such graph is shown in Fig.  \ref{Graph}, where the
system is divided into three regions,  $\mathcal{A}$,
$\mathcal{B}$ and $\mathcal{C}$. We call all edges that go from
one region to the other the \emph{boundary-crossing edges} and
label the subset of all such edges by $\mathcal{X}$. All qubits
connected by the boundary-crossing edges are in turn called the
\emph{boundary qubits} and the subset composed of all of these is called
$\mathcal{Y}$.

%%%%%%%%%%%%%%%%%%%%%%%%%%%%%%%%%%%%%%%%%%%%%%%%%%%%%%%%%%%%%%%
%%%%%%%%%%   DISCUSSING DECOHERENCE   %%%%%%%%%%%%%%%%%%%%%%%%%%

\emph{Open-system dynamics.--} Our ultimate goal is to quantify
the entanglement in any partition of arbitrary graph states undergoing a generic
physical process during a time interval $t$. The action of such
process on an initial density operator $\rho$ can be described by
a completely-positive trace-preserving map $\Lambda$ as
$\rho_t=\Lambda(\rho)$, where $\rho_t$ is the evolved density
matrix after time $t$. All such maps can be expressed in a Kraus
representation, $\Lambda(\rho)=\sum_\mu p_\mu K_\mu \rho
K_\mu^\dag$, where $\sqrt{p_\mu}K_\mu$ are called the Kraus operators (each of
which appearing with  probability $p_\mu$), which satisfy the
normalization conditions Tr$[K_\mu^\dag K_\mu]=1$ and
$\sum_\mu p_\mu =1$ \cite{nielsen}. The Kraus representation
guarantees that the map is (completely) positive  and preserves
trace normalization. When the map can be factorized as
the composition of individual maps acting independently on each
qubit, the noise is said to be individual (or independent); if
not, it is said to be collective.

A very important class of  processes is described by the Pauli
maps, separable (non-entangling) maps whose Kraus operators are
given by tensor products of Pauli operators $X$, $Y$,
$Z$, and the identity. Examples of these are the collective or
individual depolarizing, dephasing or bit-flip channels
\cite{nielsen}. As we show next, it is possible to determine the
exact entanglement evolution of graph and
graph-diagonal states (whose formal definition is provided below) subject to individual Pauli maps.

%%%%%%%%%%%%%%%%%%%%%%%%%%%%%%%%%%%%%%%%%%%%%%%%%%%%%%%%%%%%%%%%%%%%%%%
%%%%%%%%%%%%      EXACT BOUND  %%%%%%%%%%%%%%%%%%%%%%%%%%%%%%%%%%%%%%%

\emph{Exact entanglement of graph states under Pauli maps.--}  

Let
us start by recalling that a graph state is the simultaneous eigenvector -- of eigenvalue 1 --  of the $N$ generators of the
stabilizer group, that is, of the $N$ operators consisting each of which of one $X$ acting on each single qubit  and $Z$'s on all its neighboring ones \cite{graph_review}. Therefore, the
application of an $X$ or $Y$ operator on a qubit $k$ of a graph
state is equivalent to the application of $Z$ operators on all
neighboring qubits of $k$, or on all of its neighboring qubits and
on $k$ itself, respectively. The action of any Pauli map $\Lambda$
on a graph state is thus equivalent to that of another separable
map, $\tilde{\Lambda}$, whose Kraus operators $\tilde{K}_\mu$ are
obtained from $K_\mu$ replacing in the latter each $X$ and $Y$
operators by tensor products of $Z$ and identity operators
according to the rule just described \cite{comment}. 
Thus we need to consider how a general combination of $Z$ operators acts on a graph state. We use the multi-index
$\tilde{\mu}=(\mu_1,...,\mu_N)$, with $\mu_i=\{0,1\}$, to denote such a combination through $Z^{\mu_1} \otimes Z^{\mu_2}\otimes...\otimes Z^{\mu_N}$.
The action of such operator on a graph state $\ket{{G_{(\mathcal{V},\mathcal{E})}}_0}$
generates another graph state
$\ket{{G_{(\mathcal{V},\mathcal{E})}}_{\tilde{\mu}}}$, orthogonal to the
former one \cite{graph_review,HeinDurBrie}.  These considerations imply that $\rho_t$  can be expressed as 
 \begin{eqnarray}
\label{graph diagonal} \nonumber
\rho_t&=&\Lambda(\ket{{G_{(\mathcal{V},\mathcal{E})}}_0})=\tilde{\Lambda}(\ket{{G_{(\mathcal{V},\mathcal{E})}}_0})\\
&=&\sum_{\tilde{\mu}} \tilde{p}_{\tilde{\mu}}
\ket{{G_{(\mathcal{V},\mathcal{E})}}_{\tilde{\mu}}}\bra{{G_{(\mathcal{V},\mathcal{E})}}_{\tilde{\mu}}}.
\end{eqnarray}
All possible
$2^N$ graph states $\ket{{G_{(\mathcal{V},\mathcal{E})}}_{\tilde{\mu}}}$
associated to the graph $G_{(\mathcal{V},\mathcal{E})}$ form a complete
orthonormal basis of the $N$-qubit Hilbert space. State \eqref{graph diagonal} is a {\it graph-diagonal state}.
Calculating the exact entanglement in any partition of the such state is in general a problem that involves
an optimization over the entire parameter space of $\rho_t$. In
what follows we will show that it is possible to greatly reduce the complexity
of this optimization problem. Consider any partition of the state
$\rho_t$.  We now factor out explicitly all the $CZ$ gates  but those
corresponding to the boundary-crossing edges and write the state
as
\begin{eqnarray}
\label{rho_t} \rho_t&=&\bigotimes_{\{i,j\} \in \mathcal{E}/\mathcal{X}}
CZ_{ij}\sum_{\gamma,\delta} \tilde{p}_{\gamma,\delta}
\ket{{G_{(\mathcal{Y},\mathcal{X})}}_\gamma}\bra{{G_{(\mathcal{Y},\mathcal{X})}}_\gamma}\nonumber\\
&\otimes&\ket{{g_{(\mathcal{V}/\mathcal{Y})}}_\delta}\bra{{g_{(\mathcal{V}/\mathcal{Y})}}_\delta}\bigotimes_{\{k,l\} \in \mathcal{E}/\mathcal{X}}CZ_{kl},
\end{eqnarray}
%
%\begin{eqnarray}
%\label{rho_t} \rho_t&=&\bigotimes_{\{i,j\} \in \mathcal{E}/\mathcal{X}}
%CZ_{ij}\sum_{\gamma,\delta} \tilde{p}_{\gamma,\delta}
%\ket{{G_{(\mathcal{Y},\mathcal{X})}}_\gamma}\bra{{G_{(\mathcal{Y},\mathcal{X})}}_\gamma}\nonumber\\
%&\otimes&\ket{{g_{(\mathcal{V}/\mathcal{Y})}}_\delta}\bra{{g_{(\mathcal{V}/\mathcal{Y})}}_\delta},
%\end{eqnarray}
Here we have grouped together all indices inside $\tilde{\mu}$ into two new multiple indices,  $\gamma$
and $\delta$. Multiple index $\gamma$ accounts for all possible graph
states $\ket{{G_{(\mathcal{Y},\mathcal{X})}}_\gamma}$ generated by applying tensor products of $Z$ and identity
operators to the graph state
$\ket{{G_{(\mathcal{Y},\mathcal{X})}}_0}\equiv\bigotimes_{\{i,j\}
\in\mathcal{X}} CZ_{ij}\otimes\ket{{g_{(\mathcal{Y})}}_0}$,
associated to the {\it boundary graph}
$G_{(\mathcal{Y},\mathcal{X})}=\{\mathcal{Y},\mathcal{X}\}$, with
$\ket{{g_{(\mathcal{Y})}}_0}\equiv\bigotimes_{i \in\mathcal{Y}}
\ket{+_i}$. Multiple index $\delta$ on the other hand accounts for all states $\ket{{g_{(\mathcal{V}/\mathcal{Y})}}_\delta}$ 
generated from $Z$ or identity operators
on the state
$\ket{{g_{(\mathcal{V}/\mathcal{Y})}}_0}\equiv\bigotimes_{i
\in\mathcal{V}/\mathcal{Y}} \ket{+_i}$ of the non-boundary qubits
$\mathcal{V}/\mathcal{Y}$. Probability
$\tilde{p}_{\gamma,\delta}$  is defined as the sum of all $p_{\mu}$ such that
$\tilde{K}_\mu\ket{{G_{(\mathcal{Y},\mathcal{X})}}_0}\otimes
\ket{{g_{(\mathcal{V}/\mathcal{Y})}}_0}=\ket{{G_{(\mathcal{Y},\mathcal{X})}}_\gamma}\otimes\ket{{g_{(\mathcal{V}/\mathcal{Y})}}_\delta}$.
Because the $CZ$ gates explicitly factored out in state \eqref{rho_t}
%explicit $CZ$ gates in the state \eqref{rho_t} 
are
local unitary operations with respect to the partition of
interest, the entanglement of $\rho_t$, $E(\rho_t)$,
reads
 \begin{eqnarray}\label{ent}
E\big(\sum_{\gamma,\delta}
\tilde{p}_{\gamma,\delta}\ket{{G_{(\mathcal{Y},\mathcal{X})}}_\gamma}
\bra{{G_{(\mathcal{Y},\mathcal{X})}}_\gamma}\otimes\ket{{g_{(\mathcal{V}/\mathcal{Y})}}_\delta}\bra{{g_{(\mathcal{V}/\mathcal{Y})}}_\delta}\big).
 \end{eqnarray}
where $E$ is any convex entanglement quantifier not
increasing under LOCC. In what follows, we first establish a lower and
upper bound to this expression and then show that these bounds
coincide, obtaining the exact expression of the graph-state
entanglement evolution.

\par First, consider an LOCC protocol consisting of measuring all the non-boundary qubits $\mathcal{V}/\mathcal{Y}$ of the state within brackets in Eq. \eqref{ent} in the product basis composed by all orthonormal  states $\{\ket{{g_{(\mathcal{V}/\mathcal{Y})}}_\delta}\}$ and tracing out the measured subsystem after communicating the outcomes. The remaining subsystem $\mathcal{Y}$ is {\it flagged} by each measurement outcome $\delta$ -- meaning that outcome $\delta$ provides full information about to which state $\mathcal{Y}$ has been projected after each measurement run.
The final entanglement after the entire protocol is then given by
the average entanglement over all measurement runs. Since $E$ is
non-increasing under LOCC, $E(\rho_t)$ must satisfy
\begin{equation}
\label{perfectlower} E(\rho_t)\geq\sum_\delta\tilde{p}_\delta E\Big(\sum_{\gamma}
\tilde{p}_{(\gamma|\delta)}\ket{{G_{(\mathcal{Y},\mathcal{X})}}_\gamma}
\bra{{G_{(\mathcal{Y},\mathcal{X})}}_\gamma}\Big),
\end{equation}
where $\tilde{p}_{\delta}\equiv\sum_{\gamma}\tilde{p}_{\gamma,\delta}$ is the total probability of occurrence of an event $\delta$ and $\tilde{p}_{(\gamma|\delta)}$ is the conditional probability of an event $\gamma$ given that  event $\delta$ has happened.

\par On the other hand, convexity of $E$ implies that $E(\rho_t)$, as given by \eqref{ent}, must necessarily be smaller or equal to $\sum_\delta \tilde{p}_\delta E\Big(\sum_{\gamma}
\tilde{p}_{(\gamma|\delta)}\ket{{G_{(\mathcal{Y},\mathcal{X})}}_\gamma}
\bra{{G_{(\mathcal{Y},\mathcal{X})}}_\gamma}\otimes\ket{{g_{(\mathcal{V}/\mathcal{Y})}}_\delta}\bra{{g_{(\mathcal{V}/\mathcal{Y})}}_\delta}\Big)$,
% $\sum_\delta E\big(\sum_{\gamma} \tilde{p}_{\gamma,\delta}\ket{{G_{(\mathcal{Y},\mathcal{X})}}_\gamma}\bra{{G_{(\mathcal{Y},\mathcal{X})}}_\mu}\otimes$
%$\ket{{g_{(\mathcal{V}/\mathcal{Y})}}_\delta}\bra{{g_{(\mathcal{V}/\mathcal{Y})}}_\delta}\big)$,
which, since locally added ancillary systems do not change the
entanglement, is in turn equal to the right-hand side of
\eqref{perfectlower}. This means that the right-hand side of Eq.
\eqref{perfectlower} provides at the same time an upper and a
lower bound to $E(\rho_t)$ and therefore
yields its {\it exact} value, \ie:

\begin{equation}
\label{exact}
E(\rho_t)=\sum_\delta\tilde{p}_\delta E\Big(\sum_{\gamma}
\tilde{p}_{(\gamma|\delta)}\ket{{G_{(\mathcal{Y},\mathcal{X})}}_\gamma}
\bra{{G_{(\mathcal{Y},\mathcal{X})}}_\gamma}\Big).
\end{equation}

\par A comment on the implications of this exact result on the computational-cost is now in place. The calculation of the entanglement of systems composed by
$N=N_\mathcal{Y}+N_{\mathcal{V}/\mathcal{Y}}$ qubits (being
$N_\mathcal{Y}$ and $N_{\mathcal{V}/\mathcal{Y}}$ the number of
boundary and non-boundary qubits respectively) is a problem that,
in general, involves an optimization over $O\big(2^{2N}\big)$ real
parameters. Through Eq. \eqref{exact} such calculation is reduced to
that of the average entanglement over a sample of
$2^{N_{\mathcal{V}/\mathcal{Y}}}$ states  (one for each
measurement outcome $\delta$) of  $N_\mathcal{Y}$  qubits, which
involves at most $2^{N_{\mathcal{V}/\mathcal{Y}}}$ optimizations
over $O\big(2^{2N_\mathcal{Y}}\big)$ real parameters. Thus the
present method provides an exponential decrease in the computational power needed to calculate $E(\rho_t)$, since only the boundary qubits appear in the computation of Eq. \eqref{exact}.

\begin{figure}%[t]
\begin{center}
\includegraphics[width=.80\linewidth]{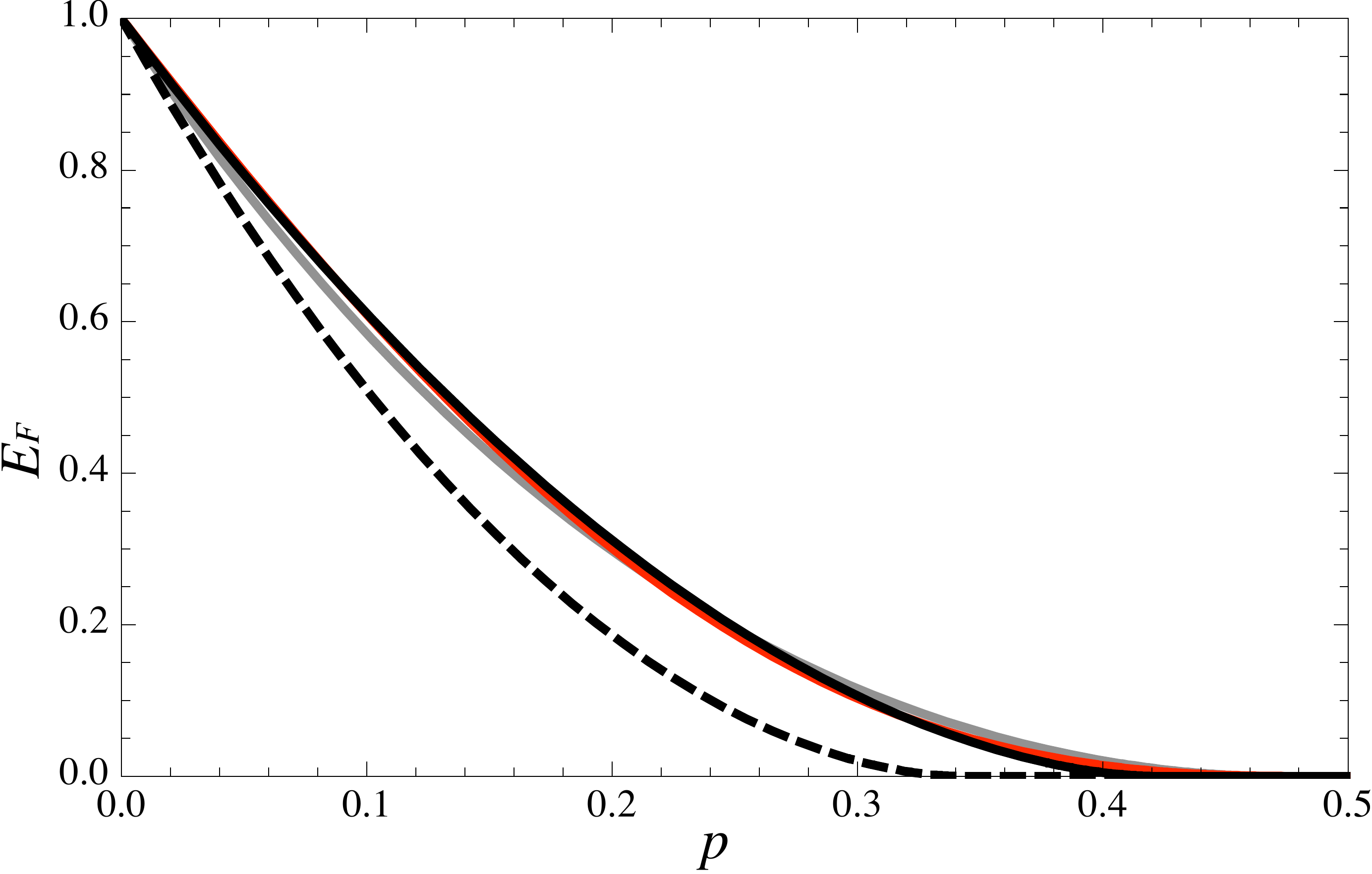}
\caption{ \label{Fig2} (Color online.) Entanglement of formation
($E_F$) in the partition of the first particle versus the rest for
1-D graph states of 2 (black), 4 (grey) and 7 (red) particles
undergoing individual depolarization as a function of the
depolarization probability $p$. The dashed curve represents a
size-independent lower bound. }
\end{center}
\end{figure}

In order to illustrate the power of the method we have calculated, using
\eqref{exact}, the exact entanglement of formation $E_F$ \cite{Wootters} of  1-D graph states  under the action of independent depolarizing channels, which  mix, with probability $p$,  any one-qubit state with the maximally mixed state $\openone/2$~ \cite{nielsen}.  In Fig. \ref{Fig2} we display the curves corresponding to the bipartition first qubit versus the rest, although other partitions can be considered. The 1-D
graph state (also called the linear cluster state), given by
$\ket{LC}=\bigotimes_{i=1}^{N-1} CZ_{i,i+1} \bigotimes_k^N\ket{+_k}$, 
evolves from $p=0$ towards a final maximally mixed state at $p=1$. Not only this calculation would have been impossible had we attempted  a brute-force optimization approach, but also, since in this particular case the boundary qubits are just two, the use of \eqref{exact} allows to perform the calculation with no optimization at all, for an explicit formula for the entanglement of formation exists for arbitrary two-qubit systems  \cite{Wootters}. 

%%%%%%%%%%%%%%%%%%%%%%%%%%%%%%%%%%%%%%%%%%%%%%%%%%%%%%%%%%%%%%%%%%%%%%%%%%%
%%%%%%% DISCUSS GENERAL INITIAL STATES AND CHANNELS%%%%%%%%%%%%%%%%%%%%%%%%

\par \emph{Beyond graph states and Pauli maps.--} The expression \eqref{exact} is actually a method for calculating
the entanglement of any graph-diagonal state as the one in \eqref{graph diagonal}.
Since Pauli maps acting on initial graph-diagonal states also produce graph-diagonal states,
all the arguments used so far are also valid for this class of initial states. Furthermore,
any quantum state can be depolarized to a graph-diagonal state by means of LOCC \cite{ADB}.
Using again the fact that the entanglement of a state does not increase if an LOCC protocol is applied,
one can see that the present method also provides (in general non-tight) lower bounds to the decay of the entanglement of any initial state
subject to any decoherence process.

%%%%%%%%%%%%%%%%%%%%%%%%%%%%%%%%%%%%%%%%%%%%%%%%%%%%%%%%%%%%%%%%%%%%%%%%%%%%%%%%
%%%%%         SIZE-INDEPENDENT LOWER BOUND   %%%%%%%%%%%%%%%%%%%%%%%%%%%%%%%%%%%

\par\emph{Robustness of graph-state entanglement.--} The developed techniques can be further simplified to obtain new lower bounds to graph-state entanglement during all the evolution that, despite not being tight, can be calculated in a much more efficient way than \eqref{perfectlower} and often turn out to be independent of the total number of qubits. This dramatically simplifies the study of the entanglement robustness of graph states as a function of the system's size, a central question for the applicability of these states as quantum information resources. 
As an illustration, we compare next graph states of different sizes under the action of general $N$-qubit Pauli maps $\Lambda$ that scale 
with $N$ in a way such that, for each $\sqrt{p_{\mu}}K_\mu$, the Kraus operators of the map acting on $M$ more qubits are obtained 
as tensor products of $\sqrt{p_{\mu}}K_\mu$ with Pauli or identity operators on the other $M$ qubits, weighted with some new probabilities that sum  up to one (for each $\mu$). That is, so that the total probability of 
event $\mu$ on the $N$ first qubits, $p_{\mu}$, remains the same. All the individual or collective Pauli maps mentioned above fall into this category.
The state between brakets in Eq. \eqref{ent} can then be written
also as $\sum_{\gamma}
\tilde{p}_{\gamma}\ket{{G_{(\mathcal{Y},\mathcal{X})}}_\gamma}\bra{{G_{(\mathcal{Y},\mathcal{X})}}_\gamma}\otimes\sum_\delta
\tilde{p}_{(\delta|\gamma)}\ket{{g_{(\mathcal{V}/\mathcal{Y})}}_\delta}\bra{{g_{(\mathcal{V}/\mathcal{Y})}}_\delta}\big)$,
where $\tilde{p}_{\gamma}\equiv\sum_{\delta}\tilde{p}_{\gamma,\delta}$ and $\tilde{p}_{(\delta|\gamma)}$ is the conditional probability of $\delta$ given  $\gamma$. By tracing out the state of
the non-boundary qubits ({\it i.e.}, by disregarding the flag that lead to \eqref{perfectlower} above)   and using again the fact that $E$ does not
increase under LOCC, we arrive at 
\begin{equation}\label{size-independent LB}
E(\rho_t)\geq E\big(\sum_{\gamma}
\tilde{p}_{\gamma}\ket{{G_{(\mathcal{Y},\mathcal{X})}}_\gamma}\bra{{G_{(\mathcal{Y},\mathcal{X})}}_\gamma}\big).
\end{equation}
Now, notice that -- for the maps here-considered --
probability $\tilde{p}_{\gamma}$ depends only on the boundary graph $G_{(\mathcal{Y},\mathcal{X})}$ and the number of non-boundary
qubits directly connected to it (the boundary graph is affected by the noise on up to its first neighbors), not on the total system size $N$. Bound  \eqref{size-independent LB} is  unaffected by the addition
of $M$ extra particles if these new particles are not connected to the boundary subsystem. In the latter sense, and for the considered noise scenario, noisy graph-state
entanglement is thus robust with respect to the variation of the system size provided  $G_{(\mathcal{Y},\mathcal{X})}$ and its connectivity to the rest do not vary. 

\par Size-independent bound \eqref{size-independent LB} (for the case of $E=E_F$) is compared with the exact entanglement, again for a linear cluster and the individual depolarizing channel, in Fig. \ref{Fig2}

%%%%%%%%%%%%%%%%%%%%%%%%%%%%%%%%%%%%%%%%%%%%%%%%%%%%%%%%%%%%%%%%%%%%%%%%%%%%%%%%%%%%%%%%
%%%%%%%%%%%%%%%%%%       DISCUSSIONS    %%%%%%%%%%%%%%%%%%%%%%%%%%%%%%%%%%%%%%%%%%%%%%

\emph{Discussion.--}
To summarize, in this work we have presented a general framework to study the entanglement decay of graph states under decoherence. It is important to emphasize that any function that satisfies the requirements of convexity and monotonicity under LOCC falls into the range of applicability of the machinery here-developed. This includes genuine multipartite entanglement quantifiers, as well as those
functions aiming at quantifying the usefulness of quantum states for given
quantum informational tasks.

\par To conclude with, let us make the following observations.
First, the techniques developed to obtain perfect bounds can also be applied to tackle some cases other than Pauli maps. For
example, for graph states in the presence of individual thermal
baths at arbitrary temperature, an LOCC procedure similar to the
one used to obtain the bound \eqref{perfectlower}, but using
general measurements instead of orthogonal ones, can be used to
obtain highly non-trivial entanglement lower bounds.

Second,
bound \eqref{size-independent LB}, when restricted to bipartite
entanglement, provides the same type of lower bound as the one used in section V-B of Ref.
\cite{HeinDurBrie} to find lower bounds to the entanglement lifetime for the case of $E$ being the negativity. The present bound has the advantage of dealing
with other possible partitions and general entanglement quantifiers. All
these topics will be touched upon 
elsewhere.

%Finally, we envision future research
%over the presented machinery that could open the way not only to the
%calculation of the exact entanglement of more general noisy states
%without optimization on the entire system's parameter space.

\begin{acknowledgements}
We thank J. Eisert, M. Plenio, F. Brand\~ao, and A. Winter, for inspiring conversations, and the CNPq,
the Brazilian Millenium Institute for Quantum Information, the PROBRAL CAPES/DAAD,
the European QAP, COMPAS and PERCENT projects, the Spanish MEC FIS2007-60182 and Consolider-Ingenio QOIT projects, and the Generalitat de Catalunya, for financial support.

\end{acknowledgements}


\begin{thebibliography}{20}
\bibitem{comment} Notice that different indices $\mu$ can often lead to the same operator $\tilde{K}_\mu$ (the same product of $Z$ operators).
\end{thebibliography}
\end{document}